\documentclass[aps,twocolumn,groupedaddress,showpacs]{revtex4}
\usepackage{graphicx}
\begin{document}
\bibliographystyle{apsrev}


\title{Phonon softening and anomalous mode near the $x_{c}=0.5$  quantum critical point in Ca$_{2-x}$Sr$_{x}$RuO$_4$}



\author{R. G. Moore$^{1}$}
\altaffiliation{Current Address: Stanford Synchrotron Radiation Laboratory, Stanford Linear Accelerator Center and
Stanford Institute for Materials and Energy Sciences, Menlo Park, CA 94025.}

\author{M. D. Lumsden$^2$}
\author{M. B. Stone$^2$}
\author{Jiandi Zhang$^3$}
\author{Y. Chen$^4$}
\author{J. W. Lynn$^4$}
\author{R. Jin$^{5,1}$}
\author{D. Mandrus$^{5,1}$}
\author{E. W. Plummer$^{1}$}

\affiliation{$^1$Department of Physics and Astronomy, The University of Tennessee, Knoxville, TN 37996}


\affiliation{$^2$Neutron Scattering Sciences Division, Oak Ridge National Laboratory, Oak Ridge, TN 32831}

\affiliation{$^3$Department of Physics, Florida International University, Miami, FL 33199}

\affiliation{$^4$NIST Center for Neutron Research, National Institute of Standards and Technology, Gaithersburg,
MD,  20899}

\affiliation{$^5$Materials Science and Technology Division, Oak Ridge National Laboratory, Oak Ridge, TN 37831}


\date{\today}

\begin{abstract}
Inelastic neutron scattering is used to measure the temperature dependent phonon dispersion in
Ca$_{2-x}$Sr$_{x}$RuO$_{4}$ ($x=0.4$, $0.6$). The in-plane \ensuremath{\Sigma}$_{4}$ octahedral tilt mode softens
significantly at the zone boundary of the high temperature tetragonal (HTT)
\textit{I4}$_{\mathit{1}}$\textit{/acd} structure as the temperature approaches the transition to a low
temperature orthorhombic (LTO) \textit{Pbca} phase. This behavior is similar to that in La$_2$CuO$_4$, but a new
inelastic feature that is not found in the cuprate is present.  An anomalous phonon mode is observed at energy
transfers greater than the \ensuremath{\Sigma}$_{4}$ albeit with similar dispersion. This anomalous phonon mode
never softens below $\sim 5$~meV, even for temperatures below the HTT-LTO transition. This mode is attributed to
the presence of intrinsic structural disorder within the \textit{I4}$_{\mathit{1}}$\textit{/acd} tetragonal
structure of the doped ruthenate.
\end{abstract}
\maketitle

The discovery of exotic superconductivity in Sr$_{2}$RuO$_{4}$ and its structural similarity to La$_{2}$CuO$_{4}$
has generated much interest in the Ca$_{2-}$$_{\mathit{x}}$Sr$_{\mathit{x}}$RuO$_{4}$ (CSRO) family of
compounds\cite{MaenoNature94,MaenoPT01,NakatsujiPRLPRB00,FriedtPRB01,NakatsujiPRL03}. Their physical properties as
a function of doping have remarkable similarities with the high-temperature superconductor,
La$_{2-}$$_{\mathit{x}}$Sr$_{\mathit{x}}$CuO$_{4}$ (LSCO) \cite{KeimerPRB92}. Nevertheless, it is important to
realize that Sr and Ca are isoelectronic, such that Ca-substitution does not change the valence electron number,
i.e. does not change the band filling in contrast with Sr doping of the
cuprate\cite{NakatsujiPRLPRB00,FriedtPRB01,KeimerPRB92,FangNJP05}.  In addition to providing a new system where
the evolution from antiferromagnetism (AFM) to superconductivity can be explored, the phase diagram of the single
layered Ruddlesden-Popper CSRO compounds contains rich and exotic behavior attributed to numerous, nearly
degenerate, structural and magnetic instabilities. For instance for $x < 0.2$, an AFM insulating ground state and
metal-insulator phase transitions are observed\cite{NakatsujiPRLPRB00,FriedtPRB01,NakatsujiPRL03}, while a
metamagnetic transition is observed for $x\sim0.2$ accompanied by anisotropic thermal expansion anomalies that can
be reversed in a magnetic field\cite{NakatsujiPRLPRB00,BaierPB06}.  For $0.2 \leq x < 0.5$, short range AFM
correlations exist but vanish at $x_{c} \approx 0.5$ where the spin susceptibility is critically enhanced
indicating a ferromagnetic instability point\cite{NakatsujiPRLPRB00,FriedtPRB01,FangNJP05}. This critical
concentration, $x_{c} \approx 0.5$, is also the $T = 0$~K terminus \textit{quantum critical point} (QCP) for the
line of structural phase transitions between the HTT and LTO phases\cite{NakatsujiPRLPRB00,FriedtPRB01}.

\begin{figure}
\includegraphics[keepaspectratio=true, width =
3 in]{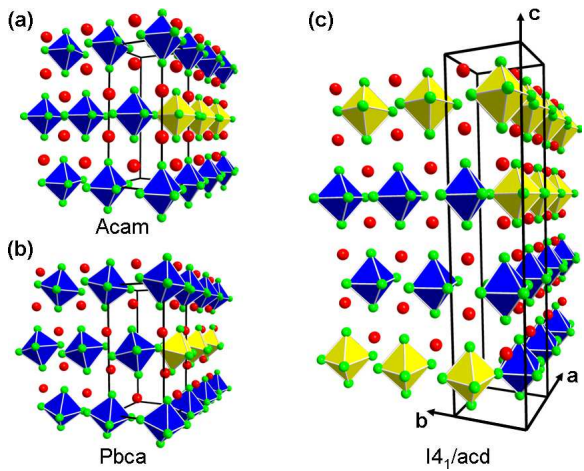} \caption{Structural phases of the CSRO family for $x \leq 1.5$. RuO$_6$ octahedra
are illustrated with yellow (blue) octahedra representing clockwise (counter-clockwise) rotation relative to the
$c$-axis. Oxygen (Ca/Sr) sites are represented by green (red) spheres and unit cells are shown as black lines.
Only octahedra along the [100] and [010] faces are shown for clarity.  CSRO $x \leq 0.2$ presents the
\textit{Acam} symmetry (a) at high temperatures and \textit{Pbca} symmetry (b) at low temperatures. Both $x=0.4$
and $x=0.6$ present the \textit{I4}$_{\mathit{1}}$\textit{/acd} symmetry (c) at high temperatures but the LTO
phase of $x=0.4$ involves a structural frustration between the preferred RuO$_6$ rotation stacking periodicity of
the \textit{I4}$_{\mathit{1}}$\textit{/acd} phase and the induced tilt periodicity of the \textit{Pbca} phase
\cite{FriedtPRB01,FriedtThesis}.} \label{StackingSymmetry}
\end{figure}

Inelastic neutron scattering (INS) measurements of the temperature dependence of the \ensuremath{\Sigma}$_{4}$
phonon mode in Ca$_{2-}$$_{\mathit{x}}$Sr$_{\mathit{x}}$RuO$_{4}$ ($x = 0.4$, $0.6$) were performed. These
concentrations are in immediate proximity to the QCP while allowing investigation of mode softening both with and
without traversing the HTT-LTO phase boundary. This choice of concentrations is also motivated by interest in
surface phases and phase transitions in CSRO, and the possibility of resolving questions concerning surface
mechanisms by understanding the bulk phonon behavior\cite{Surface214Group}.

Figure~\ref{StackingSymmetry} summarizes the three structural phases of the CSRO compounds for $x \leq 0.5$.
Generically, the structure consists of Ru sites in an octahedral coordination with neighboring O sites.  The
octahedra form a layered structure in the $ab$-plane with neighboring planes along the $c$-axis separated by Ca/Sr
layers. The structural phase transitions are associated with changes of the octahedral tilt and rotation as well
as the octahedral stacking sequence along the $c$-axis.  While both $x=0.4$ and $x=0.6$ start in the HTT
\textit{I4}$_{\mathit{1}}$\textit{/acd} symmetry at room temperature, $x=0.4$ enters into a LTO \textit{Pbca}
symmetry upon cooling\cite{FriedtPRB01,NakatsujiPRLPRB00,FriedtThesis}.  While the LTO phase is created by a
static tilt of the RuO$_6$, structural frustration is introduced due to the preferred stacking periodicities along
the $c$-axis observed for $x$ away from the QCP\cite{FriedtThesis}. The \ensuremath{\Sigma}$_{4}$ transverse
acoustic phonon mode is related to the tilt of the layered Ru octahedra and, as such, is particularly sensitive to
their distortions in different portions of the CSRO phase diagram.

Single crystal samples were grown using the floating zone technique, $m_{x=0.4} \approx 3$~g and $m_{x=0.6}
\approx 4$~g.  INS measurements for $x=0.4$ were performed using the HB1 triple-axis spectrometer (TAS) at the
High Flux Isotope Reactor (HFIR) at ORNL and the BT-7 TAS at the NIST Center for Neutron Research. INS
measurements for $x=0.6$ were performed using the HB3 TAS at HFIR.  The HB1 and HB3 instrument configurations
consisted of a fixed focus PG(002) monochromator and a flat PG(002) analyzer with collimations 48'-40'-40'-240'.
The BT-7 configuration consisted of a variable-focus PG(002) monochromator and a focusing PG(002) analyzer with
collimations open-50'-40'-open. All measurements employed a fixed final neutron energy of $14.7$~meV with a PG
filter in the scattered beam. Samples were mounted in the (hhl) scattering plane and indexed in the
\textit{I4/mmm} symmetry of pure Sr$_{2}$RuO$_{4}$. Unless noted, all (hkl) coordinates refer to this notation.
The (1.5 1.5 2) wave vector is a zone boundary for both the \textit{I4/mmm} and
\textit{I4}$_{\mathit{1}}$\textit{/acd} symmetries. However, due to the static RuO$_{6}$ rotation, a glide plane
symmetry is established and the (1.5 1.5 2) Bragg peak is extinguished for $T>T_{C}$. As one cools through the
HTT-LTO transition, the ensuing static distortion results in the appearance of this Bragg peak as this wave vector
becomes a zone center for the orthorhombic \textit{Pbca} phase. Using the Bragg peak intensity as an order
parameter, we find $T_{c} \sim 155$~K for the $x=0.4$ sample, consistent with prior
studies\cite{NakatsujiPRLPRB00,FriedtPRB01}.  No hysteresis for the order parameter is observed, consistent with
the expected second order phase transition.

Constant wave vector scans were performed to determine the temperature dependent dispersion of the
\ensuremath{\Sigma}$_{4}$ phonon mode. This phonon propagates in the [1 1 0] direction, but the motion of the
oxygen atoms due to the static rotation results in mixed longitudinal and transverse components. Regions near the
(1.5 1.5 2) and (0.5 0.5 6) wave vectors provided clean $\Sigma_4$ phonon measurements because of structure factor
considerations similar to those of La$_{2}$CuO$_{4}$ and Sr$_{2}$RuO$_{4}$~\cite{BirgeneauPRL87,BradenPRB98}.

\begin{figure}
\includegraphics[keepaspectratio=true, width = 2.4 in] {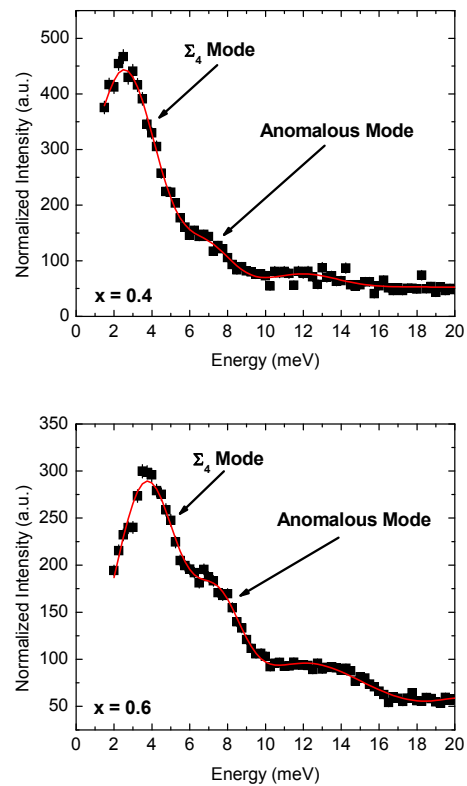}
\caption{Zone boundary (1.5 1.5 2) INS data for $x=0.4$ at $T = 270$~K and $x=0.6$ at $T = 200$~K. The
\ensuremath{\Sigma}$_{4}$ and anomalous modes are annotated and observed in all scans.  Red line is the result
from multiple Gaussian fits.  Uncertainties where indicated are statistical and represent one standard deviation.}
\label{ZoneBoundaryRaw}
\end{figure}

\begin{figure}
\includegraphics[keepaspectratio=true, width =
3.4 in]{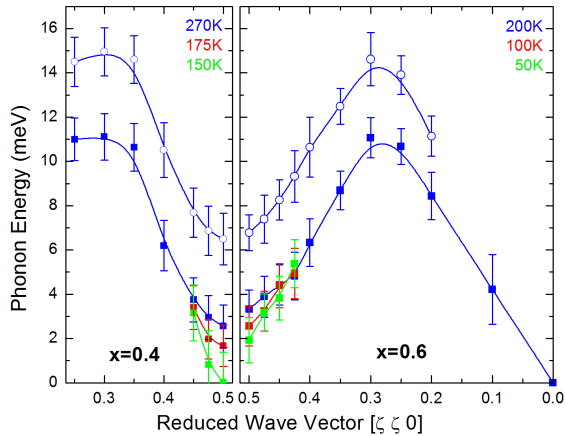} \caption{Dispersion of the \ensuremath{\Sigma}$_{4}$ (filled symbols) and
anomalous (open symbols) phonon modes for both $x=0.4$ and $x=0.6$.  The dispersion curves are plotted with the
$\it{I4/mmm}$ Brillouin Zone of the parent compound Sr$_2$RuO$_4$.  The \ensuremath{\Sigma} $_{4}$ mode shows
typical soft mode behavior and the anomalous mode mimics the \ensuremath{\Sigma}$_{4}$ dispersion.  Lines are
added as guides to the eye.} \label{DispersionSummary}
\end{figure}

\begin{figure}
\includegraphics[keepaspectratio=true, width = 3.2 in]{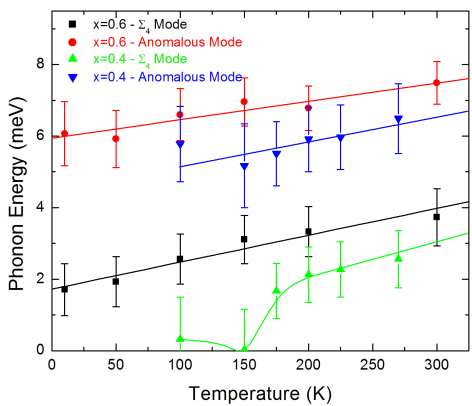} \caption{Zone boundary softening for the \ensuremath{\Sigma}$_{4}$ and anomalous modes. The \ensuremath{\Sigma}$_{4}$ mode reaches zero energy as the phase boundary is crossed.}
\label{ZoneBoundaryTemp}
\end{figure}

Phonon dispersion curves were found by comparing individual constant wave vector scans to multiple Gaussian peaks.
A Gaussian peak was assumed for the incoherent peak plus one for each mode observed.  All data were corrected for
monitor contamination due to higher order Bragg scattering from the PG(002) monochromator.  Data collected with a
fixed focus monochromator were corrected to account for deviations in beam size as a function of incident energy.
To improve the comparison and allow all fit parameters to be unrestricted during analysis, the data were first
smoothed using wavelet shrinkage \cite{KolaczykAJ97, CharlesSIA04}. Fig.~\ref{ZoneBoundaryRaw} shows constant-Q
scans, corrected as described above, at the (1.5 1.5 2) Brillouin zone boundary for $x=0.4$ and $x=0.6$.

The scans shown in Fig.~\ref{ZoneBoundaryRaw} reveal several spectral features with the lowest energy feature
being the \ensuremath{\Sigma}$_{4}$ mode. In addition, a slightly higher energy phonon mode is observed in both
samples. This mode appears in all Brillouin zones where the \ensuremath{\Sigma}$_{4}$ mode is present and
possesses an intensity and wave-vector modulation which mimics the \ensuremath{\Sigma}$_{4}$ mode. This anomalous
mode is not expected from normal mode analysis ~\cite{BradenPRB98,FriedtThesis} and, as such, it is anomalous in
nature. The fundamental reasons for the existence of this anomalous mode must be explored to fully understand the
physics of the system.  In addition to these two modes, a higher energy lattice excitation is also seen in
Fig.~\ref{ZoneBoundaryRaw} that is not examined here.

Constant-Q scans near (1.5 1.5 2) and (0 0 6) are combined to form the dispersion curves shown in
Fig.~\ref{DispersionSummary}. This dispersion is plotted in the \textit{I4/mmm} notation to allow comparison with
similar soft mode behavior in La$_{2}$CuO$_{4}$ \cite{BirgeneauPRL87}.  The anomalous mode is clearly visible in
Fig.~\ref{DispersionSummary} with a dispersion that mimics the \ensuremath{\Sigma}$_{4}$ mode displaced in energy.
This is in contrast to La$_{2}$CuO$_{4}$ where only the single \ensuremath{\Sigma}$_{4}$ mode is observed. The
\ensuremath{\Sigma}$_{4}$ phonon mode is doubly degenerate with two possible tilt modes around [0$\bar{1}$0] and
[100]. The simplest explanation for the presence of two modes would be a lifting of this degeneracy.  However,
since the space group is tetragonal, the [0$\bar{1}$0] and [100] directions are identical and we don't expect the
degeneracy to be lifted~\cite{BirgeneauPRL87,BradenPRB98}. Cooling results in increased softening of the
\ensuremath{\Sigma}$_{4}$ mode in both samples.  For $x=0.6$, the phonon energy never reaches zero at the zone
boundary as the crystal remains in the tetragonal phase.  A summary of the temperature dependent softening of the
two modes at (1.5 1.5 2) is shown in Fig.~\ref{ZoneBoundaryTemp}. Similar linear softening with temperature is
observed in both samples for both modes with the exception of a deviation from linearity for $x=0.4$ below $200$~K
as the lower branch softens to zero energy at $T\sim150$~K.  Note that while this mode softens completely, its
corresponding anomalous mode never softens below $5$~meV.

SrTiO$_{3}$ and La$_{2}$CuO$_{4}$ are classic examples where soft phonon behavior drives structural instabilities
~\cite{ShiraneRMP74,BirgeneauPRL87}. Displacive phase transitions are typically associated with a soft phonon mode
that freezes into a static lattice distortion at a critical temperature\cite{CowleyAP80}. For example, the energy
of the \ensuremath{\Sigma}$_{4}$ tilt mode reduces to zero at the Brillouin zone boundary in La$_{2}$CuO$_{4}$ at
the HTT-LTO phase transition \cite{BirgeneauPRL87}. The $\Sigma_{4}$ transverse acoustic phonon mode represents a
rotation (in-plane tilt) of the CuO$_6$ octahedron about an axis in the $ab$-plane\cite{GrandeAC77}. Because
Ca$_{2-x}$Sr$_{x}$RuO$_{4}$ has the same oxygen octahedron structure and also undergoes a HTT-LTO phase transition
($0.2 < x < 0.5$), one expects similar softening behavior in the CSRO family. However, there are significant
differences in symmetry for the LSCO and CSRO systems. At higher values of $x \approx 1.5$, CSRO has already
undergone a structural transition resulting from the freezing of a $\Sigma_{3}$ phonon mode with corresponding
static rotation of the octahedra about the $c$-axis~\cite{FriedtPRB01}. Thus, the HTT-LTO transition in CSRO
(LSCO) is from space group \textit{I4}$_{\mathit{1}}$\textit{/acd} to \textit{Pbca} (\textit{I4/mmm} to
\textit{Cmca}). One obvious consequence of the different space groups is that there is no change in the shape of
the Brillouin zone during the HTT-LTO phase transition in CSRO.  Despite this difference, the point group symmetry
for Ru is identical in both \textit{I4/mmm} and \textit{I4}$_{\mathit{1}}$\textit{/acd}, and therefore, one does
not expect any degeneracy lifting as a result of this difference.

Little is known about the tetragonal-to-tetragonal phase transition in CSRO that occurs at $x \approx1.5$. The
transition is from the space group \textit{I4/mmm} to \textit{I4}$_{\mathit{1}}$\textit{/acd}, caused by the
rotation of the RuO$_{6}$ octahedra about the $c$-axis. It is believed that the $x \approx1.5$ structural
transition in Sr$_{2}$RuO$_{4}$ results from the softening of the $\Sigma_{3}$ mode\cite{BradenPRB98}. For $0.5
\leq x < 1.5$, a second order structural phase transition into the \textit{I4}$_{\mathit{1}}$\textit{/acd} phase
is observed\cite{FriedtPRB01,FriedtThesis}.  Although the symmetry is \textit{I4}$_{\mathit{1}}$\textit{/acd},
disorder in the $c$-axis periodicity of the RuO$_6$ rotation is observed. Such disorder introduces stacking faults
resulting in a mixture of \textit{I4}$_{\mathit{1}}$\textit{/acd} and \textit{Acam} symmetries as shown in
Fig.~\ref{StackingSymmetry}.  As $x$ is decreased, the disorder in the $c$-axis periodicity is reduced and a more
perfect \textit{I4}$_{\mathit{1}}$\textit{/acd} phase is formed \cite{FriedtPRB01,FriedtThesis}.

While Friedt tentatively attributes the anomalous mode to interactions resulting from the different stacking
periods of the tilt and rotational distortions combined with disorder in Ca/Sr mixing\cite{FriedtThesis}, the
static energy of the anomalous mode as the $\Sigma_{4}$ mode softens through the phase boundary suggests an
alternate mechanism. We propose a simple model of intrinsic disorder within the
\textit{I4}$_{\mathit{1}}$\textit{/acd} symmetry to explain the anomalous mode. While the symmetry for $0.5 < x <
1.5$ is \textit{I4}$_{\mathit{1}}$\textit{/acd}, disorder along the $c$-axis exists with a coherence length of
approximately two unit cells observed for $x\sim1.0$ \cite{FriedtThesis}.  It has also been observed that as more
Ca is added to the system, a more perfect \textit{I4}$_{\mathit{1}}$\textit{/acd} symmetry forms resulting in
fewer faults and an increased coherence length.  It should be noted that long range order exists within individual
$ab$-planes and it is only the $c$-axis disorder in the stacked layers of rotated RuO$_{6}$ that varies with $x$.
Disorder in the $c$-axis octahedral rotation periodicity has also been observed in similar materials such as
Sr$_{2}$IrO$_{4}$ and Sr$_{2}$RhO$_{4}$ ~\cite{DisorderGroup}. While inter-layer coupling along the $c$-axis is
assumed weak as previous experiments suggest, it must occur for the well coordinated RuO$_6$ rotations and tilts
to exist along the $c$-axis~\cite{BradenPRB98,FriedtPRB01}. The \textit{I4}$_{\mathit{1}}$\textit{/acd} symmetry
encodes four RuO$_{6}$ layers with a $c$-axis lattice parameter $\sim 25$ \AA.  The structural frustration created
from the rotational periodicity mismatch combined with the intrinsic stacking faults could lead to the formation
of impurity domains of different symmetry (\textit{Acam}) along the $c$-axis and lift the phonon degeneracy. While
locally the domains would appear to be \textit{Acam}, faults are isolated and random and the lack of $c$-axis
correlation does not present a mixed phase scenario nor would it appear in x-ray or neutron diffraction
measurements. However, since the phonon propagates in the $ab$-plane, local disorder could lift the degeneracy
allowing for spectral intensity in inelastic measurements. It has been observed in 2D systems that disorder can
alter phonon softening and prevent the freezing phonon from reaching zero energy at the phase
transition\cite{PetersenPSS02}. If the anomalous mode is due to disorder in the quasi 2D layered system then one
could expect that such a mode would soften to a finite temperature at the phase boundary as the
\ensuremath{\Sigma}$_{4}$ mode reduces to zero energy.

In summary, inelastic neutron scattering experiments have been performed to measure the dispersion of the
\ensuremath{\Sigma}$_{4}$ tilt mode phonon in Ca$_{1.4}$Sr$_{0.6}$RuO$_{4}$. The \ensuremath{\Sigma}$_{4}$ mode
demonstrates typical soft phonon behavior similar to La$_{2}$CuO$_{4}$, but a new anomalous phonon mode appears.
The anomalous mode mimics the \ensuremath{\Sigma}$_{4}$ dispersion except at the phase boundary where the
anomalous mode remains at finite energy while the $\Sigma_{4}$ mode softens to zero energy creating the
orthorhombic phase.  The anomalous mode is most likely due to disorder in the layered stacking sequence lifting
the \ensuremath{\Sigma}$_{4}$ degeneracy.  Further investigation is required to fully understand the role of
defects and stacking faults in the CSRO family.


\begin{acknowledgments}
We thank I. A. Sergienko for helpful discussions.  This work is supported by NSF DMR-0346826, DMR-0353108,
DMR-0451163, DOE DE-FG02-04ER46125, DOE DMS, and ORAU faculty summer research program. A portion of this research
at Oak Ridge National Laboratory's High Flux Isotope Reactor was sponsored by the Scientific User Facilities
Division, Office of Basic Energy Sciences, DOE. The work at Oak Ridge National Laboratory was supported through
the Division of Materials Sciences and Engineering, Office of Basic Energy Sciences, DOE under Contract
DE-AC05-00OR22725.
\end{acknowledgments}


%
%

%
%

\end{document}